\documentclass[screen,authorversion]{acmart}
\AtBeginDocument{%
 }

\copyrightyear{2025}
\acmYear{2025}
\setcopyright{rightsretained}
\acmConference[LAK 2025]{LAK25: The 15th International Learning Analytics and Knowledge Conference}{March 03--07, 2025}{Dublin, Ireland}
\acmBooktitle{LAK25: The 15th International Learning Analytics and Knowledge Conference (LAK 2025), March 03--07, 2025, Dublin, Ireland}
\acmDOI{10.1145/3706468.3706530}
\acmISBN{979-8-4007-0701-8/25/03}

\usepackage{xcolor}
\definecolor{mygreen}{RGB}{0, 150, 0} 
 \usepackage{multirow}
 
\usepackage{tabularx}
\usepackage{multirow}

\begin{document}

\title[Does Multiple Choice Have a Future in the Age of Generative AI?]{Does Multiple Choice Have a Future in the Age of Generative AI? A Posttest-only RCT}

\author{Danielle R. Thomas}
\email{drthomas@cmu.edu}
\affiliation{
 \institution{Carnegie Mellon University}
 \city{Pittsburgh, PA}
 \country{USA}
}

\author{Conrad Borchers}
\email{cborcher@cs.cmu.edu}
\affiliation{%
 \institution{Carnegie Mellon University}
 \city{Pittsburgh, PA}
 \country{USA}
}

\author{Sanjit Kakarla}
\email{sanjit.kakarla@gmail.com}
\affiliation{%
 \institution{Carnegie Mellon University}
 \city{Pittsburgh, PA}
 \country{USA}
}

\author{Jionghao Lin}
\email{jionghao@cmu.edu}
\affiliation{%
 \institution{Carnegie Mellon University}
 \city{Pittsburgh, PA}
 \country{USA}
}

\author{Shambhavi Bhushan}
\email{shambhab@andrew.cmu.edu}
\affiliation{%
 \institution{Carnegie Mellon University}
 \city{Pittsburgh, PA}
 \country{USA}
}

\author{Boyuan Guo}
\email{boyuan@andrew.cmu.edu}
\affiliation{%
 \institution{Carnegie Mellon University}
 \city{Pittsburgh, PA}
 \country{USA}
}

\author{Erin Gatz}
\email{egatz@andrew.cmu.edu}
\affiliation{%
 \institution{Carnegie Mellon University}
 \city{Pittsburgh, PA}
 \country{USA}
 }
 
\author{Kenneth R. Koedinger}
\email{koedinger@cmu.edu}
\affiliation{
 \institution{Carnegie Mellon University}
 \city{Pittsburgh, PA}
 \country{USA}
}

\renewcommand{\shortauthors}{Thomas et al.}

\begin{abstract}
The role of multiple-choice questions (MCQs) as effective learning tools has been debated in past research. While MCQs are widely used due to their ease in grading, open response questions are increasingly used for instruction, given advances in large language models (LLMs) for automated grading. This study evaluates MCQs effectiveness relative to open-response questions, both individually and in combination, on learning. These activities are embedded within six tutor lessons on advocacy. Using a posttest-only randomized control design, we compare the performance of 234 tutors (790 lesson completions) across three conditions: MCQ only, open response only, and a combination of both. We find no significant learning differences across conditions at posttest, but tutors in the MCQ condition took significantly less time to complete instruction. These findings suggest that MCQs are as effective, and more efficient, than open response tasks for learning when practice time is limited. To further enhance efficiency, we autograded open responses using GPT-4o and GPT-4-turbo. GPT models demonstrate proficiency for purposes of low-stakes assessment, though further research is needed for broader use. This study contributes a dataset of lesson log data, human annotation rubrics, and LLM prompts to promote transparency and reproducibility.

\end{abstract}

\ccsdesc[500]{Human-centered computing~Human computer interaction (HCI)}
\ccsdesc[500]{Applied computing~Computer-managed instruction}
\ccsdesc[500]{Computing methodologies~Artificial intelligence}

\keywords{Tutoring, Generative AI, Human-AI tutoring, AI-assisted tutoring, Assessment}

\maketitle

\section{Introduction}
The effectiveness of multiple-choice questions (MCQs) in learning is the subject of much debate \cite{butler2018multiple, gurung2024multiple, haladyna2004developing}. Although MCQs are often criticized for lack of depth, they remain a common feature in K-12 and higher education, due to their ease of grading \cite{butler2018multiple}. However, their potential as instructional tools, rather than just assessment tools, meaning that they provide feedback from which students can learn, has received less attention. In contrast, open-response questions are frequently used in assignments such as homework, under the assumption that they promote deeper learning \cite{butler2018multiple, gurung2024multiple}. However, open responses can be more time consuming for learners and resource intensive to grade \cite{butler2018multiple}, although recent advancements in the field have made the automated grading of these responses more feasible. This study evaluates the effectiveness of MCQs in relation to open-response questions, both individually and in combination, as learning-by-doing activities. These learning-by-doing activities are embedded in six tutoring lessons that involve advocacy training. To investigate scalability of autograding open-ended responses, we use generative AI to evaluate tutors' open responses. The contributions of this work are twofold: \textit{theoretically}, it offers insights into the learning benefits of MCQs compared to open responses in learning-by-doing instruction; and \textit{practically}, it provides implications for optimizing tutor training by determining the most efficient method of instructing and assessing tutors, as measured in the completion time of instruction and accuracy of automated open-response grading. Furthermore, this study contributes a dataset of lesson log data, human annotation rubrics, and AI-generated generative commands to improve transparency, reproducibility, and collaboration within the learning analytics community.

The design of instructional materials plays a critical role in promoting effective learning. MCQs are often favored for their efficiency--they can be administered and graded quickly and require less time for learners to respond, making them appealing in large-scale settings \cite{butler2018multiple, gurung2024multiple, haladyna2004developing}. However, MCQs are sometimes criticized for promoting surface-level learning, as they may encourage guessing and recognition rather than deeper understanding \cite{butler2018multiple}. In contrast, open-response questions require tutors to construct their answers, thereby engaging in higher-order thinking and reflection. Although open-response questions can be powerful pedagogically, they are also more time-consuming to complete and evaluate \cite{butler2018multiple}. A key question is whether MCQs can be designed to be as pedagogically effective as open-response questions within the context of teaching tutors advocacy skills, particularly in scenarios where scalability is a concern \cite{gurung2024multiple}. Advocacy, an emerging area of instruction aimed at improving equity and inclusion in tutoring, is particularly suited for a comparison of MCQ to open-ended responses, because the skills it requires--such as critical thinking and ethical reasoning--are potentially more effective for comprehension when practiced through open-ended formats \cite{butler2018multiple} rather than MCQs where generating distractors can pose a challenge \cite{dutulescu2024beyond}. Compared to STEM learning, where many closed-form grading systems exist (e.g., tutoring systems), there is a need in learning analytics to study what forms of instruction are effective in novel, less structured domains like advocacy training.

Generative AI, particularly large language models (LLMs), have the capability to evaluate tutors' textual, open responses in real-time. LLMs such as GPT-4 \cite{openai2024gpt4technicalreport}, Claude \cite{caruccio2024claude}, and LLaMa \cite{touvron2023llama} have demonstrated remarkable performance in a variety of linguistic tasks. These modern LLMs are built on a large-scale transformer architecture and trained on extensive datasets \cite{brown2020language, gallegos2024bias}. As a result, LLMs have attracted substantial interest from researchers across various fields, including education, because of their potential to perform reasoning tasks at scale and with reduced costs. Generative AI systems can evaluate human tutor responses across a wide range of scenarios, providing feedback and assessment at a scale that would be impossible for human evaluators alone. Importantly, LLMs may have the potential to make situational judgments, assessing not only the correctness of a response but also underlying reasoning \cite{brown2020language}. This capability is crucial in scenario-based training, where tutors must navigate complex real-world situations. However, despite their potential, LLMs also have limitations, such as the tendency to generate nonsensical or factually incorrect outputs \cite{zhang2023siren}, and bias and fairness issues \cite{gallegos2024bias}. Using generative AI for tutor evaluation, this study explores the potential of AI to support large-scale, effective tutor training, ultimately improving tutor quality and accessibility. 

This work addresses the need for effective and scalable tutor training by evaluating tutors' posttest performance across six scenario-based lessons focused on advocacy skills. Using a posttest-only randomized experimental design, we analyze tutor performance and time spent across three learning conditions: MCQ only, open-response questions only, and a combination of both. We evaluated the scalability of using generative AI to evaluate tutor responses, comparing the performance of GPT-4-turbo and GPT-4o with human graders. This study addresses the following research questions:

\textbf{RQ1}: What differences exist in tutor learning, as evidenced by posttest performance, across the learning-by-doing activities, i.e., MCQ only, open-response questions only, or both? 

\textbf{RQ2}: In what contexts do MCQs, open-response questions, or a combination of both yield the highest accuracy and efficiency, thereby optimizing the impact of the lesson?  

\textbf{RQ3}: How effective are LLMs, namely GPT-4o and GPT-4-turbo, in assessing tutors' open responses at posttest?

\section{Related Work}
\subsection{Tutor Advocacy Skills and Scenario-based Training}

Tutoring is widely recognized as one of the most effective interventions for improving student learning outcomes \cite{guryan2021not, nickow2020impressive, pane2014effectiveness}. Research consistently shows that personalized support from skilled human tutors can significantly boost student academic performance, particularly among struggling students \cite{robinson2021high}. However, ensuring access to adequately trained tutors is challenging \cite{chhabra2022evaluation, kraft2021blueprint}, with many tutoring organizations relying on paraprofessionals. Many paraprofessional tutors have college education, but lack formal training and  in providing instruction and building quality relationships with students \cite{chhabra2022evaluation, nickow2020impressive}. In addition, very limited instructional materials are available for tutors on attending to students' social-emotional needs. The process of training human tutors presents substantial scalability challenges, such as the need for human evaluators to assess tutor performance. Traditional methods of tutor training and evaluation are both time-consuming and resource-intensive, limiting the ability to scale tutoring programs to meet the needs of all students. Tutoring is more effective when delivered by teachers or well-trained professional tutors \cite{nickow2020impressive}. Currently, limited instructional materials are available to prepare and provide situational experiences to inexperienced tutors.

The lessons draw from previous research that identified impactful competencies of effective tutoring within the area of Advocacy \cite{chhabra2022evaluation, thomas2023tutor, chine2022scenario}. Past studies have internally validated the construct validity by demonstrating ~20\% learning gain from pretest to posttest on similarly-structured lessons covering topics related to: giving effective praise to students; reacting when a student makes an error; and determining what students know \cite{thomas2023tutor}. Using the same scenario-based structure as in previous work \cite{chine2022development, thomas2023tutor, chine2022scenario, thomas2024learning}, our goal is to optimize tutor learning focusing on tutor lessons that instruct tutors in advocacy skills. There are very limited instructional and training materials available to tutors in the area of advocacy. Advocacy in teaching and tutoring encompasses a range of skills that promote student success by addressing their academic, social-emotional, and equity-related needs \cite{chhabra2022evaluation, cook2020student}. Key areas include: promoting equity and inclusion; fostering cultural awareness; and challenging unconscious bias and assumptions \cite{thomas2023tutor, chhabra2022evaluation}.

\subsection{Learning Engineering and Instructional Design}The design of instructional materials is essential for effective tutor training, with multiple-choice questions (MCQs) often preferred for their efficiency in large-scale settings, although they may promote surface-level learning \cite{butler2018multiple, gurung2024multiple, haladyna2004developing}. In contrast, open-response questions foster deeper reflection and higher-order thinking, but are more time-consuming, raising the question of whether MCQs can be made pedagogically effective to teach advocacy skills in scalable scenarios \cite{butler2018multiple, gurung2024multiple}. This present work applies a learning engineering approach to investigate the learning efficiency of the following type of questions: open response, which encourages deeper cognitive engagement; MCQs which can provide structured assessment and objective grading; and a combination leveraging the strengths of both \cite{butler2018multiple, gurung2024multiple}. Central to this present work is the ``learn-by-doing'' methodology, which emphasizes active participation in the learning process. This approach aligns with ``doer'' philosophy, advocating for hands-on, practical experiences to enhance understanding and retention \cite{koedinger2015learning}. An example in practice is the integration of computer-based Cognitive Tutors, whereby students are required not only to complete tasks but also to articulate (analogous to the ability of tutors to explain in this current work) their reasoning, reinforcing their comprehension and retention \cite{aleven2002effective}. This dual emphasis on \textit{doing} and \textit{explaining} has been shown to significantly improve learning outcomes, fostering deeper comprehension and critical thinking skills \cite{aleven2002effective, thomas2023tutor}. 

\begin{figure*}[ht]
\centering
\hspace*{-.7cm}
\includegraphics[width=\textwidth]{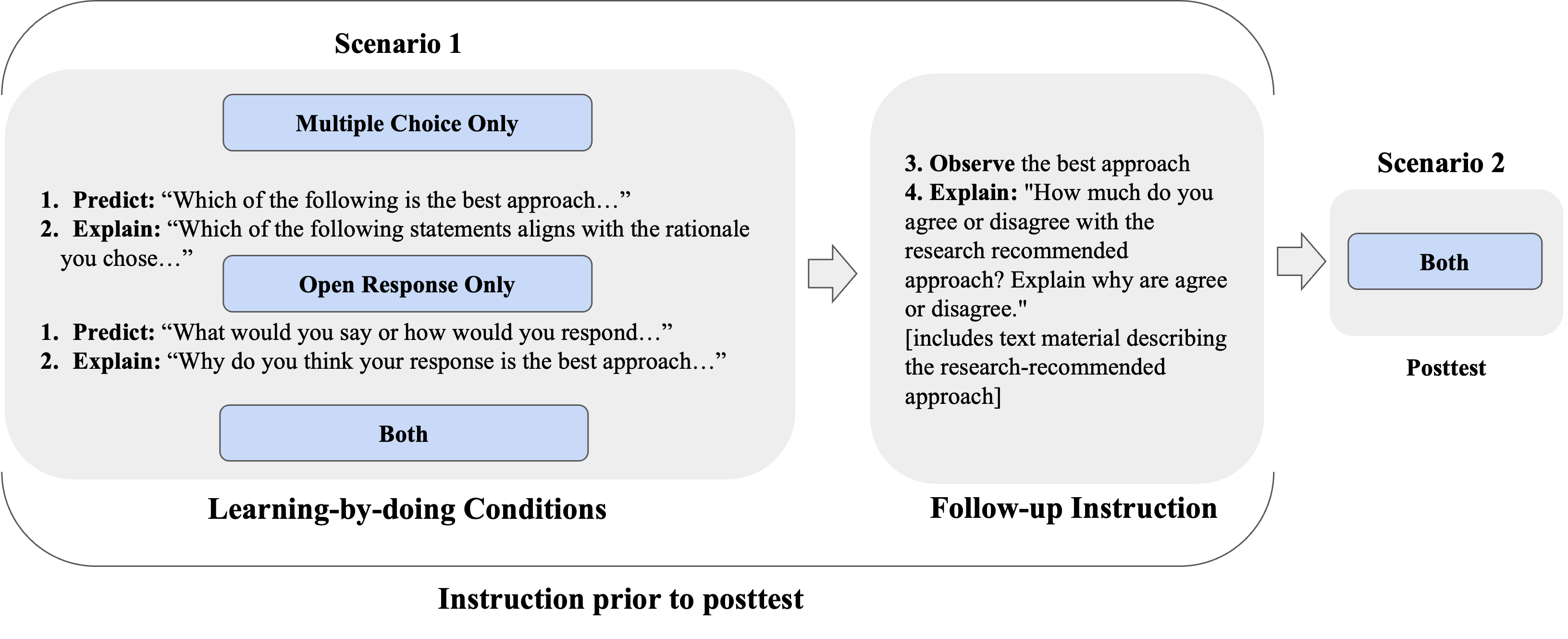}
\caption{Instructional design sequence of the lessons illustrating the three learning-by-doing conditions, then the follow-up instruction phase, and concluding with posttest. } \label{fig:1_unnamed-6}
\Description{Instructional design sequence of the lessons illustrating the three learning-by-doing conditions, then the follow-up instruction phase, and concluding with posttest. }
\end{figure*}

Brief scenario-based lessons were strategically designed using the learning-by-doing approach that provides actionable feedback and requires tutors to apply what they learned within the learning-by-doing conditions and the instruction phase by applying their learning in analogous tutoring situations at posttest. Fig. \ref{fig:1_unnamed-6} illustrates the instructional design of the lesson. First, the tutors are presented with a scenario (Scenario 1), whereby they are prompted to \textit{predict} the best approach, followed by being asked to \textit{explain} their rationale or reasoning behind their chosen approach. There are three possible learning-by-doing conditions: multiple choice only, open response only, or both. Multiple choice questions begin with \textit{``Which of the following…,''} followed by four options for the tutor to choose. Open response questions start with \textit{ ``What would you say or how would you respond…''} for predicting the best approach and \textit{``Why do you think your response is the best approach…''} for explaining the rationale behind their chosen approach. The tutors then engage in the instruction phase where the tutors observe the research-recommended approach and explain their reasoning in support or not of the best approach. Finally, the tutors complete a posttest, which is the same for all tutors and uses both MCQs and open responses. This instructional design is considered a modified predict-observe-explain (POE) approach and is theoretically related to Gibbs' Reflective Cycle, a cyclical instructional model that provides structure for learning by doing to individual learning experiences \cite{gibbs1988learning}.

\subsection{Using Generative AI to Evaluate Responses}
Traditional automated assessment methods, such as natural language processing and traditional machine learning, often struggle to capture the complexity of open-ended responses due to their limited ability to understand context, subtle language variations, and complex semantics \cite{condor2021automatic,lin2023using}. Additionally, these models require extensive labeled training data to perform accurately, which can be time-consuming and costly to gather, especially in specialized domains like education. Automated assessment of open-ended responses is a significant task in learning analytics, as it can evaluate the quality of learners' understanding of various topics, such as student responses to college readiness \cite{condor2021automatic}. Previous studies have used language models such as BERT \cite{lin2023using} and Sentence-BERT \cite{condor2021automatic} to develop automated evaluation systems for open-ended responses. Although these models have shown promising results, they have limitations in their ability to understand deeper contextual nuances and adapt to complex domains. These machine-learning models often struggle to fully capture the nuanced linguistic and semantic meanings in tutor responses. 

In response to the limitations of traditional machine learning methods, recent advancements in generative AI, particularly LLMs, have demonstrated significant potential to assess open-ended responses. For example, \cite{lee2024applying} applied GPT-3.5 and GPT-4 to automatically score student-written responses in science assessments. They found that few-shot learning approaches significantly improved scoring accuracy, especially when paired with contextual item stems and scoring rubrics. Their study demonstrated that GPT-4 outperformed GPT-3.5, highlighting the potential of generative AI to provide more accurate and explainable automatic scores in educational contexts. \cite{lin2024can} used few-shot prompting strategies with GPT-4 to assess the correctness of the responses of novice tutors in various tutoring strategies, such as giving praise, responding to student errors, and understanding student knowledge levels. \cite{yun2024enhancing} used GPT-4-turbo to evaluate the understanding of novice tutors of essential tutoring practices, including encouraging active student learning and fostering a respectful community. These studies illustrate the growing capacity of generative AI to offer more nuanced and contextually aware assessments of open-ended educational responses. 

\subsection{Prompt Engineering} Prompt engineering is a crucial technique in using LLMs to produce more accurate and contextually relevant output. The goal is to provide better context and structure in prompts to guide the LLMs toward the desired responses. Prompt engineering is particularly important when nuanced understanding and reliability are essential. One key technique in prompt engineering is few-shot prompting \cite{brown2020language}, where a set of exemplars is provided in the input prompt to demonstrate the ideal model behavior. By showing the model examples of the desired output format, few-shot prompting helps the LLM generalize to similar but unseen tasks. This approach has been used effectively in various automated assessment tasks, such as evaluating tutoring practice \cite{hanimproving, kakarla2024using} and student explanation on computer science questions \cite{carpenter-etal-2024-assessing}. Another approach is chain-of-thought (CoT) prompting \cite{wei2022chain}, which instructs the model to ``think step by step'' by outlining intermediate reasoning steps. This technique helps the model handle more complex tasks by breaking the problem down into manageable parts, leading to more coherent and accurate responses. Such techniques are also used effectively to assess responses in educational settings \cite{hirunyasiri2023comparative, lin2024can}.

In addition to structuring prompts, techniques such as self-consistency \cite{wang2022self} enhance the reliability of the model's outputs. Rather than modifying the prompt itself, self-consistency involves sampling multiple outputs and using a majority vote to determine the final response. This approach reduces the likelihood of hallucinations and improves the reliability of the outputs generated. For example, when evaluating open-ended responses, employing self-consistency can ensure that the model consistently interprets nuanced answers in a similar manner across multiple runs. Furthermore, prompting for rationale can help in obtaining more interpretable outputs. By explicitly asking the model to explain its reasoning or decision-making process, researchers can gain insight into how the model arrived at its conclusions, thus improving the transparency and interpretability of the assessment. This is particularly valuable in educational contexts, where understanding the reasoning behind a student's answer is as important as the answer itself. Overall, prompt engineering strategies such as few-shot prompting \cite{brown2020language}, chain-of-thought prompting \cite{wei2022chain}, and self-consistency \cite{wang2022self} play a vital role in enhancing the performance and reliability of LLMs in educational assessments. By carefully designing prompts and leveraging these techniques, we can guide LLMs to provide more accurate, nuanced, and consistent evaluations of open-ended educational responses.

\section{Method}
Six scenario-based lessons were created and designed to align with the tutoring competencies within the area of Advocacy \cite{chhabra2022evaluation}. The lesson content taken from the tutoring platform and formatted as documents can be found in \href{https://github.com/CMU-PLUS/LAK2025-Advocacy}{Digital Appendix}. The lesson titles and learning objectives are listed below. 
\begin{itemize}
  \item \textit{{Addressing Microaggressions}}: define the term microaggression; identify microaggressions that occur in tutoring settings; and apply equity-focused strategies to help students address microaggressions. 
  \item \textit{{Avoiding Unconscious Assumptions}}: identify unconscious assumptions; and apply strategies to prevent making unconscious assumptions while tutoring.
  \item \textit{{Building Cultural Competence}}: identify when students have different cultural backgrounds and experiences than your own; practice strategies to build cultural competence, supporting and engaging students across cultures.
  \item \textit{{Exploring Implicit Bias}}: identify implicit, or conscious bias; and apply strategies to counter the effects of your own implicit biases.
  \item \textit{{Narrowing Opportunity Gaps}}: define the term opportunity gap; identify examples of opportunity gaps in tutoring settings; and explain strategies to narrow opportunity gaps in tutoring settings.
  \item \textit{{Helping Students Manage Inequity}}: recognize when a student is experiencing inequity related to their learning; and apply strategies to help students manage inequities by assisting students to advocate for themselves.
\end{itemize}

\subsection{Tutor Participants \& Lesson Delivery}There were 234 tutors, mainly college students, who completed any number of the six lessons for a total of 790 lesson completions. The tutors were undergraduate college students employed as paid tutors for a remote tutoring organization supporting middle-school students. While the demographics of the tutors were undisclosed, they exhibited cultural and racial diversity. Before starting a lesson, participants provided their levels of self-reported tutoring experience using a 5-point Likert scale with 1 indicating little to no experience (novice) and 5 indicating an expert tutor. On average, the tutors reported an experience level of 3.56 (\textit{SD} = 1.09) in all lessons. Tutors also self-reported their perceived knowledge of the lesson topic using a similar Likert scale with 1 (little to no knowledge of the topic) and 5 (expert level knowledge). In all lessons, the tutors reported an average knowledge of lesson topics of 3.59 (\textit{SD} = 1.00). Table 1 displays the number of participants by condition with the average self-reported knowledge of the topic for each lesson. The advocacy lessons were developed in collaboration with tutoring supervisors, who are responsible for training tutors and working with students, and a university research team specializing in learning science, thereby enhancing construct validity. The lessons were delivered via an online tutoring platform and align with the research-shown competencies of effective tutoring \cite{chhabra2022evaluation, thomas2023tutor}. For broad dissemination and use across varying tutoring organizations, all lessons have a Flesch-Kincaid readability index measure that ranges from grades 6-9 (defined as spanning from \textit{easy to read} to \textit{average reading level}). This ensures that all tutors, regardless of their individual reading level, can understand the content of the lesson. Each participant was randomly assigned one of three conditions in the learning-by-doing phase of the lesson: \textit{multiple choice only}, \textit{open response only}, or \textit{both}. All participants received one of two randomly chosen scenarios used in the posttest for counterbalancing of scenario difficulty.

\begin{table*}[ht]
\centering
\caption{Number of tutors by lesson for each condition with the average self-reported knowledge of the lesson topic (1-5).}
\resizebox{0.98\textwidth}{!}{%
\begin{tabular}{l c c|c c|c c}
\hline
\multirow{2}{*}{Lesson} & \multicolumn{1}{l}{\multirow{2}{*}{\begin{tabular}[c]{@{}c@{}}MCQ only\\ (n)\end{tabular}}} & \multicolumn{1}{l}{\multirow{2}{*}{\begin{tabular}[c]{@{}c@{}}Topic \\ Knowledge\end{tabular}}} & \multirow{2}{*}{\begin{tabular}[c]{@{}c@{}}Open response only \\ (n)\end{tabular}} & \multicolumn{1}{l}{\multirow{2}{*}{\begin{tabular}[c]{@{}c@{}}Topic \\ Knowledge\end{tabular}}} & \multirow{2}{*}{\begin{tabular}[c]{@{}c@{}}Both \\ (n)\end{tabular}} & \multicolumn{1}{l}{\multirow{2}{*}{\begin{tabular}[c]{@{}c@{}}Topic \\ Knowledge\end{tabular}}} \\
                      & \multicolumn{1}{l}{}                                    & \multicolumn{1}{l}{}                  &                     & \multicolumn{1}{l}{}                  &              & \multicolumn{1}{l}{}                  \\ \hline
\textit{Addressing Microaggressions}    & 39                                             & 3.55                          & 42                   & 3.42                          & 41            & 3.97                          \\
\textit{Avoiding Unconscious Assumptions}  & 34                                             & 3.68                          & 37                   & 4.07                          & 39            & 3.49                          \\
\textit{Building Cultural Competence}    & 41                                             & 3.86                          & 28                   & 3.84                          & 51            & 3.42                          \\
\textit{Exploring Implicit Bias}      & 38                                             & 3.79                          & 41                   & 3.54                          & 39            & 3.66                          \\
\textit{Helping Students Manage Inequities} & 72                                             & 3.32                          & 71                   & 3.49                          & 82            & 3.37                          \\
\textit{Narrowing Opportunity Gaps}     & 37                                             & 3.73                          & 30                   & 3.68                          & 28            & 3.64                          \\ \hline
\end{tabular}%
}
\end{table*}

\subsection{Human Open-Response Annotation and Inter-rater Reliability}
Two experienced researchers scored participant responses to assess inter-rater reliability. Open-response questions tasking tutors to predict and explain the best approach were binary-coded. Correct responses (score = 1) align with the research-recommended approach of the lesson. Conversely, incorrect responses (score = 0) do not align with research-driven tutoring best practices. Human annotation rubrics for scoring tutor responses to \textit{predict} and \textit{explain} the best approach, along with learner-sourced examples of coded responses are located within the \href{https://github.com/CMU-PLUS/LAK2025-Advocacy}{Digital Appendix}. Two experienced researchers scored tutor responses to assess inter-rater reliability. Table 2 presents the inter-rater reliability between two experienced researchers for each lesson, assessing both predict and explain responses. 
\begin{table*}[ht]
\caption{Inter-rater reliability between human evaluators for each lesson.}
\resizebox{0.98\textwidth}{!}{
\begin{tabular}{lcccc}
\hline
\multirow{3}{*}{Lesson}           & \multicolumn{2}{c}{\multirow{2}{*}{\textit{Predict responses}}} & \multicolumn{2}{c}{\multirow{2}{*}{\textit{Explain responses}}} \\
                      & \multicolumn{2}{c}{}                      & \multicolumn{2}{c}{}                      \\ \cline{2-5} 
                      & Agreement (\%)         & Cohen’s Kappa ($\kappa$)        & Agreement (\%)         & Cohen’s Kappa ($\kappa$)         \\ \hline
\textit{Addressing Microaggressions}    & 85.1\%             & 0.70             & 94.7\%             & 0.89             \\
\textit{Avoiding Unconscious Assumptions}  & 94.4\%             & 0.88             & 94.4\%             & 0.87             \\
\textit{Building Cultural Competence}    & 93.8\%             & 0.73             & 87.8\%             & 0.72             \\
\textit{Exploring Implicit Bias}      & 90.3\%             & 0.55             & 84.0\%             & 0.68             \\
\textit{Helping Students Manage Inequities} & 96.7\%             & 0.86             & 96.5\%             & 0.93             \\
\textit{Narrowing Opportunity Gaps}     & 93.7\%             & 0.86             & 96.2\%             & 0.92   \\ \hline          
\end{tabular}
}
\end{table*}

Tutor performance on the open-response questions ranged from 45\% to 89\% on the predict questions and 51\% to 68\% for the explain question types. To gain perspective on lesson and question difficulty, Table 3 displays the percentage of correct responses for each lesson by question type. Overall, tutor performance was lower on questions prompting tutors to explain the rationale behind their predictions of the best approach.
\begin{table}[ht]
\caption{Percentage of correct open responses for each lesson broken out by predict and explain responses.}
\begin{tabular}{lll}
\hline
\multirow{2}{*}{Lesson}           & \multicolumn{2}{c}{\multirow{2}{*}{Correct (\%)}} \\
                      & \multicolumn{2}{c}{}                       \\ \cline{2-3} 
                      & \textit{Predict}   & \textit{Explain}   \\ \hline
\textit{Addressing Microaggressions}    & 45.5\%             & 51.8\%            \\
\textit{Avoiding Unconscious Assumptions}  & 63.3\%             & 68.5\%            \\
\textit{Building Cultural Competence}    & 87.6\%             & 67.4\%            \\
\textit{Exploring Implicit Bias}      & 89.3\%             & 56.4\%            \\
\textit{Helping Students Manage Inequities} & 86.4\%             & 55.6\%            \\
\textit{Narrowing Opportunity Gaps}     & 63.3\%             & 58.2\%      \\\hline      
\end{tabular}
\end{table}

\subsection{Prompting Large Language Models}
Drawing on prior research in prompt engineering and large language models \cite{brown2020language, lin2024using}, we developed a method for utilizing LLMs to evaluate the correctness of open-ended responses at posttest. Specifically, we implemented a few-shot learning approach, which has been shown to enhance performance in natural language understanding tasks \cite{brown2020language}. In this method, the model was provided with a set of learner-generated responses along with human-scored examples to help it generate accurate assessments. The prompts were designed to assess two distinct types of responses: (1) ``predict'' responses, where the participants predicted the best course of action, and (2) ``explain'' responses, where they justified their decisions. Tables \ref{tab:tab4} and \ref{tab:tab5} present the specific prompts used to evaluate the predict and explain responses in the \textit{Addressing Microaggressions lesson}, respectively. Prompts for all lessons and question types (i.e., predict and explain) are located within the \href{https://github.com/CMU-PLUS/LAK2025-Advocacy}{Digital Appendix}. 

The development of these prompts was iterative, with multiple rounds of refinement informed by feedback on initial outputs. We employed various prompt engineering strategies including: content framing by providing context-specific details (e.g., \textit{``You are a tutor evaluator…''}); randomly selecting human-coded examples of correct, partially correct, and incorrect tutor responses; and prompting the model to give rationale through chain-of-thought prompting \cite{wei2022chain}. To ensure deterministic outputs, the model's temperature was set to 0, guiding it to select the most probable word (or token) based on the input, thereby minimizing randomness and producing consistent, often more conservative responses. The output length was limited to 300 tokens to avoid unnecessary verbosity. These strategies allowed us to create an efficient and accurate method for assessing textual responses, with absolute performance metrics presented in the \textit{Results}. The process also raised critical considerations for deploying LLMs in assessment contexts, which are further discussed in the \textit{Discussion}.

\begin{table*}[ht]
\caption{The LLM prompt for scoring \textit{predict} responses from \textit{Addressing Microaggressions}.}
\label{tab:tab4}
\resizebox{0.98\textwidth}{!}{%
\renewcommand{\arraystretch}{1.5} %
\fontsize{11}{13}\selectfont %
\begin{tabular}{p{18cm}} %
\hline
\small
\begin{tabular}[c]{@{}p{18cm}@{}} \texttt{SCORING\_PROMPT\_START} = """\\ You are a tutor evaluator. Please score the following tutor response to a tutor training scenario involving a middle school student facing a microaggression as follows:

-if the tutor’s response acknowledges a student facing a microaggression by providing support to the student, by giving the student space to talk, or facilitating an apology, score with a 1. Sample responses scoring a 1 are "Nori, I heard what your teacher said about the exit ticket. It's okay to find problems challenging. Let's work on it together during our session. Your efforts are important, and I'm here to support you." and "Nori, I noticed what your teacher said to you could have been upsetting and I am here if you would like to talk about it."; and "Ashley, that was not a kind or respectful thing to say. Please apologize to your classmate."; "I would acknowledge Ashley's mean comment and tell her what she did wasn't right"; "Nori, I notice that you look sad on your face. Is it because what your teacher just said? Do you want to talk about it with me?"

-if the tutor’s response does not acknowledge the microaggression or does not provide any support to the student, score with a 0. Yes or no questions, if they don't acknowledge the student's experience are scored with a 0. Sample responses scoring a 0 are "Nothing is too difficult if you decide you want to do it Nori. You are capable"; "Do you want to talk about what the teacher said to you"; "Nori, I heard what your teacher said to you about the exit ticket. Do you want to discuss it with me"; and "I would address the problem."

Response Start {-}{-}{-}"""\\ \texttt{FORMAT\_PROMPT} = \\ "{-}{-}{-} Response End. Given the earlier transcript, please return a JSON string following the format, \{\textbackslash{}"Rationale\textbackslash{}": \textbackslash{}"your reasoning here\textbackslash{}", \textbackslash{}"Score\textbackslash{}":0/1\}."\end{tabular}
\\ \hline 
\end{tabular}
}
\end{table*}

\begin{table*}[ht]
\caption{The LLM prompt for scoring \textit{explain} responses from \textit{Addressing Microaggressions}.}
\label{tab:tab5}
\resizebox{0.98\textwidth}{!}{%
\renewcommand{\arraystretch}{1.5} %
\fontsize{11}{13}\selectfont %
\begin{tabular}{p{18cm}} %
\hline
\small
\begin{tabular}[c]{@{}p{18cm}@{}} \texttt{SCORING\_PROMPT\_START} = """\\ You are a tutor evaluator. Please assess a tutor’s response within a tutor training scenario involving a tutor instructing a middle school student who has faced a microaggression: The tutor is explaining the rationale behind their response. Assess and score the tutor’s response, as follows:

-if the tutor's response demonstrates that they understand how to recognize and acknowledge a microaggression by providing the student support or issuing an apology, score with a 1. Sample responses scoring a 1 include: "Acknowledging the student's feelings and naming the microaggression, the teacher's comment, will provide an opportunity to address the microaggression"; "This approach will acknowledge the microaggression because it directly addresses Nori's feelings and opens up a supportive dialogue. By acknowledging that the teacher's comment may have been hurtful, it validates Nori's experience and gives her the opportunity to express her emotions."

-if the tutor's response does not demonstrate that the tutor recognizes how they should acknowledge microaggressions, score with a 0. Sample responses scoring a 0 include: "Telling the student she is capable of solving the problem will boost her confidence and addressing the problem will help to boost the students emotional status"; "It encourages the student to work on the problem"; and "This will provide her with a safe space to communicate."

Response Start {-}{-}{-}"""\\ \texttt{FORMAT\_PROMPT} = \\ "{-}{-}{-} Response End. Given the earlier transcript, please return a JSON string following the format, \{\textbackslash{}"Rationale\textbackslash{}": \textbackslash{}"your reasoning here\textbackslash{}", \textbackslash{}"Score\textbackslash{}":0/1\}."\end{tabular} \\ \hline 
\end{tabular}
}
\end{table*}
\subsection{Research Design \& Analysis Plan}
We employed a posttest-only randomized experimental design to evaluate tutor performance across three distinct conditions within the learning-by-doing phase: multiple-choice questions only, open-response questions only, or a combination of both. Participants were randomly assigned to one of these conditions to ensure that any differences observed in the posttest outcomes could be attributed to the specific question format rather than pre-existing differences among participants. A posttest-only design was chosen to avoid potential biases that might arise from pretesting, such as testing effects or sensitization \cite{trochim2001research}. Assessing tutor performance solely on the posttest scenario, we aimed to obtain a clear measure of the impact of the different learning-by-doing conditions on the tutor’s subsequent performance \cite{trochim2001research}. Attending to \textbf{RQ1}, we employed an ANOVA to examine the impact of lesson, conditions, and scenario order on tutor performance, with all factors being between subjects. Regarding \textbf{RQ2}, we replicated the ANOVA used for RQ1 on the time it took students to complete the learning-by-doing and follow-up instruction in each condition. For answering \textbf{RQ3}, we used prompt engineering of large language models, GPT-4 and GPT-4o, to evaluate tutors' open responses. Then employed the same ANOVA model from RQ1 to determine if the results are synonymous with the analysis of human-graded tutor responses. We then report the absolute performance of both LLM models.  

\subsubsection{Lesson Log Data}
Student responses to individual practice questions, survey questions, and other forms of instruction (e.g., responses to multiple-choice options, Likert scales, and open-ended responses) were recorded in PSLC DataShop, an open repository for educational log data commonly used for tutoring systems in learning analytics research \cite{koedinger2015learning}. Specifically, data was recorded in transaction format, which means that we analyzed individual interactions of students with timestamps. We prioritized maintaining the privacy and confidentiality of tutors, adhering to all Institutional Review Board (IRB) requirements. The lesson log data can be accessed within the \href{https://github.com/CMU-PLUS/LAK2025-Advocacy}{Digital Appendix}.  

To measure performance on the posttest (RQ1), we aggregated the accuracy of student responses on the posttest, where students completed two multiple-choice and two open-response questions, with the latter graded by two experienced researchers and LLM models (RQ3). To measure the time students took to complete the instruction (RQ2), we calculated the difference between the instruction start time, as recorded in log data, and the last student response to questions associated with each lesson’s instruction. In cases where students completed lessons over multiple sessions with substantial breaks between them, we excluded the breaks from the total lesson completion time. Due to the right-skewed distribution of completion times, which are always greater than zero, we applied a logarithmic transformation for ANOVA and statistical tests that assume normally distributed outcomes. This assumption was confirmed in the logarithmic transformation data through visual inspection of standard diagnostic plots (e.g., residual Q-Q plots). However, for ease of interpretation, we re-transformed the averages and confidence intervals from the log scale back to the standard time scale (i.e., minutes) for presentation in plots.

\section{Results}
\subsection{\textbf{Learner Performance Across Conditions}} 

\begin{figure*}[ht]
\includegraphics[width=\textwidth]{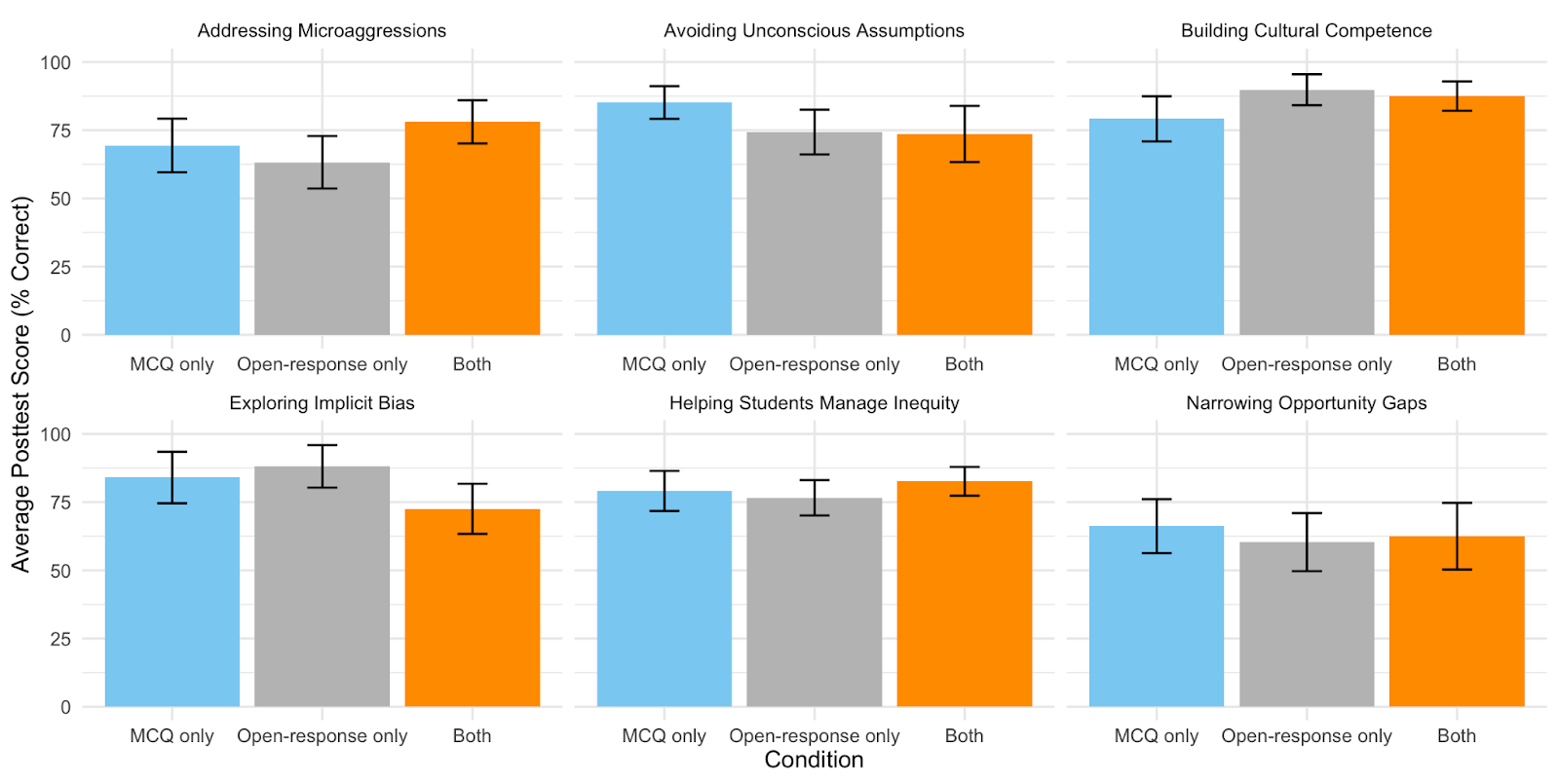}
 
 \caption{Average posttest scores compared across learning-by-doing conditions: \textit{MCQ Only}, \textit{Open-response Only}, or \textit{Both}. No significant differences were found in posttest scores between conditions. Error bars represent 95\% confidence intervals.}
 \label{fig:1_unnamed-8}
 \Description{Average posttest scores compared across learning-by-doing conditions: MCQ Only, Open-response Only, or Both. No significant differences were found in posttest scores between conditions. Error bars represent 95\% confidence intervals.}
\end{figure*}

The overall average results of the posttest are shown in Fig. \ref{fig:1_unnamed-8}. As is visually apparent, there is no consistent pattern illustrating that one instructional condition produces better learning outcomes than others across the conditions. Indeed, in an analysis of variance, we did not find a statistically significant main effect of condition, \emph{F}(2, 717) = 0.27, \emph{p} = .765. There was a significant interaction between condition and lesson \emph{F} (10, 717) = 2.20, \emph{p} =.012, which means that the posttest scores differed significantly by condition depending on what lesson the tutors completed. Furthermore, a significant main effect of the lesson suggested substantial accuracy differences by lesson, which means that some lessons were harder than others, on average, \emph{F}(5, 717) = 10.18, \emph{p} < .001.

We did \textit{post hoc} contrasts using marginal means to estimate which condition-level differences within lessons were reliable \cite{emmeans}. Significant differences were found in only two lessons. In the \textit{Addressing Microaggressions} lesson, the \textit{Both} condition produced higher posttest scores than the \textit{Open-response Only} condition (estimate = 0.136, SE = 0.054, $p = .031$) with the \textit{MCQ Only} condition ambiguously in between (i.e., not statistically different from the other two conditions, $p$-values $> 0.12$). The results for the \textit{Exploring Implicit Bias} lesson were essentially the opposite, consistent with the overall interaction. In the \textit{Exploring Implicit Bias} lesson, the \textit{Both} condition produced lower posttest scores than the \textit{Open-response Only} condition (estimate = -0.175, SE = 0.054, $p = .004$). In this lesson, the \textit{Both} condition also produced lower posttest scores than the \textit{MCQ Only} condition (estimate = -0.134, SE = 0.056, $p = .046$). All other comparisons were not significant ($p > .299$).

\subsection{\textbf{RQ2: Optimizing Lesson Impact }}
\begin{figure*}[ht]
\includegraphics[width=\textwidth,keepaspectratio]{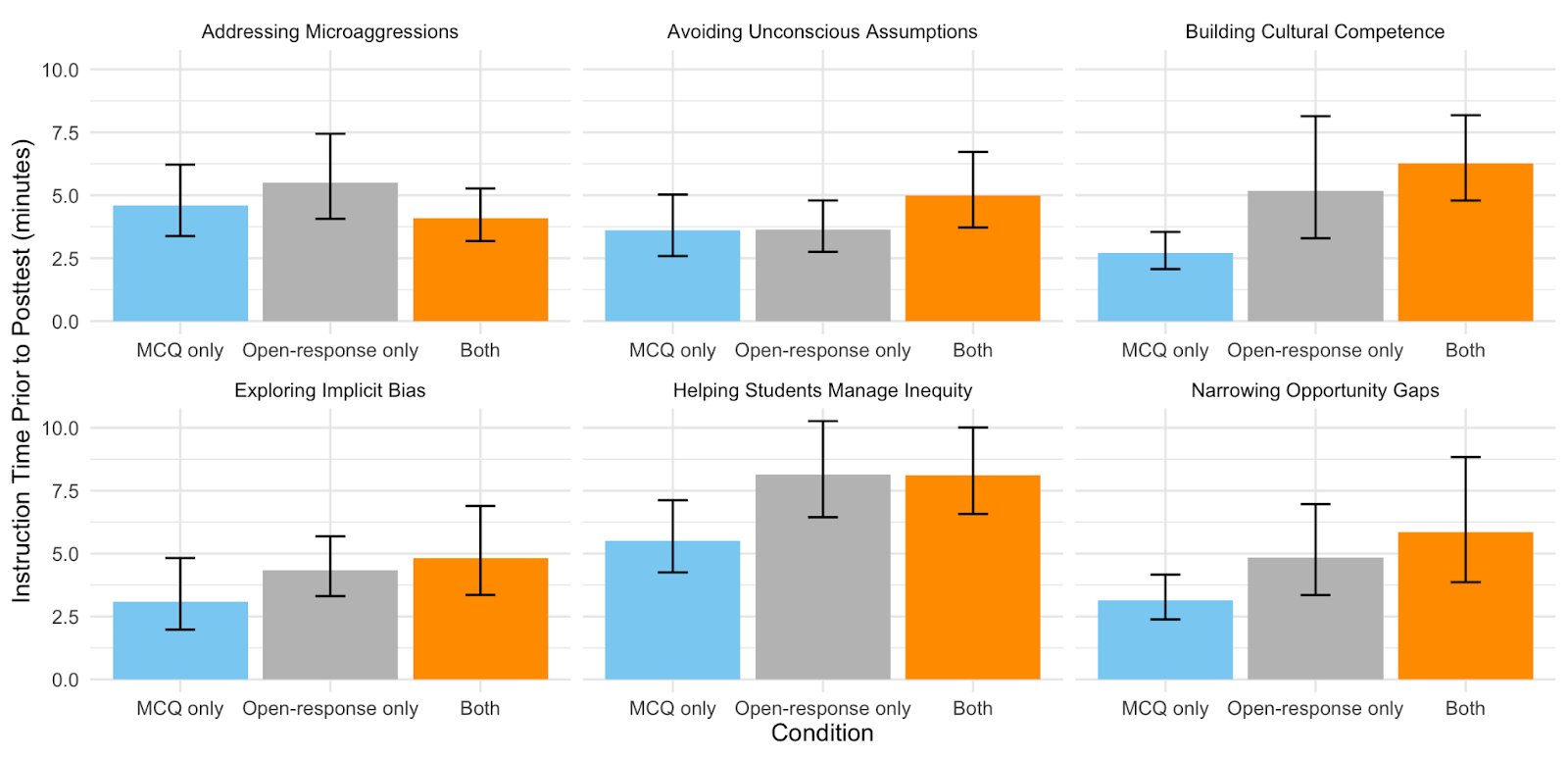}

 \caption{Average instruction time prior to posttest compared across learning-by-doing conditions: \textit{MCQ Only, Open-response Only, or \textit{Both}}. Although \textit{MCQ Only} took less time on average, no overall significant differences in instruction time prior to posttest were found between the conditions. Error bars represent 95\% confidence intervals.}
 \label{fig:1_unnamed-9}
 \Description{Average instruction time prior to posttest compared across learning-by-doing conditions: MCQ Only, Open-response Only, or Both. Although MCQ Only took less time on average, no overall significant differences in instruction time prior to posttest were found between the conditions. Error bars represent 95\% confidence intervals.}
\end{figure*}

There was no significant interaction between condition and lesson when comparing \textit{instruction time prior to posttest} (See Fig. \ref{fig:1_unnamed-6}), \emph{F}(10, 716) = 13.46, \emph{p} = .199, indicating that the time taken did not significantly differ by condition depending on the lesson students completed. However, there were significant main effects of both condition and lesson. A significant main effect of condition suggested that the time taken differed substantially by condition, \emph{F}(2, 716) = 12.56, \emph{p} < .001, while a significant main effect of lesson indicated that some lessons took longer to complete than others, on average, \emph{F}(5, 716) = 8.61, \emph{p} < .001. Across lessons, total instruction time prior to posttest ranged between \emph{M} = 2.65 minutes (\emph{Avoiding Unconscious Assumptions}) to \emph{M} = 5.60 minutes (\emph{Helping Students Manage Inequity}) (Fig. \ref{fig:1_unnamed-9}). Further, the \emph{MCQ Only} condition took students the shortest (\emph{M} = 3.83 minutes) while the \emph{Both} condition took them the longest time (\emph{M} = 5.87 minutes), although not substantially longer than the \emph{Open-response Only} condition (\emph{M} = 5.38 minutes). Hence, the \emph{Both} condition took students less time than the sum of the open and MCQ conditions. Based on marginal mean comparisons, these differences were statistically significant, such that the \emph{MCQ Only} condition took learners significantly shorter than the \emph{Both} condition, estimate = –0.37, \emph{SE} = 0.07, \emph{p} < .001, and the \emph{Open-response Only} condition, estimate = -0.30, \emph{SE} = 0.07, \emph{p} < .001. However, there was no significant difference in completion time between the \emph{Open-response Only} and \emph{Both} conditions, estimate = -0.07, \emph{SE} = 0.07, \emph{p} = .642.

Although the overall interaction between condition and lesson was not significant, we report marginal mean contrasts of conditions within lessons similar to RQ1. The \emph{MCQ Only} condition took significantly less time than the \emph{Both} condition in the \emph{Building Cultural Competence} and \emph{Narrowing Opportunity Gaps} (estimate = -0.55, \emph{SE} = 0.19, \emph{p} = .011). In the \emph{Building Cultural Competence} lesson, \emph{MCQ Only} was also significantly faster than the \emph{Open-response Only} (estimate = -0.50, \emph{SE} = 0.19, \emph{p} = .025) and \emph{Both} conditions (estimate = -0.76, \emph{SE} = 0.16, \emph{p} < .001). Notably, in no lesson was the \emph{Open-response Only} condition significantly different from the \emph{Both} condition (\emph{p}-values > .269). Overall, these findings suggest that the \emph{Both} condition took learners significantly longer than the \emph{MCQ Only} condition in 2/6 lessons, but not longer than the \emph{Open-response Only} condition.

\subsection{\textbf{Large Language Model Performance}} The absolute performance of the LLMs was determined in assessing tutors’ open response at posttest. Tables 6 and 7 display the comparison of absolute performance for GPT-4o and GPT-4-turbo, respectively, across lessons for \emph{predict} and \emph{explain} open responses. The LLM prompts used for each lesson can be accessed in the \href{https://github.com/CMU-PLUS/LAK2025-Advocacy}{Digital Appendix}. Both GPT-4o and GPT-4-turbo showcased proficiency in evaluating tutors’ responses. Accuracy measures the proportion of correct predictions out of all predictions, providing an overall sense of a model's performance. $AUC$ (Area Under the Curve), often used with imbalanced datasets, assesses how well the model distinguishes between classes, while the $F_1$ score balances precision and recall, offering a measure of the model's effectiveness in handling both false positives and false negatives. In general, both models demonstrated proficiency in performance, with accuracy ranging from 71\% to 91\% for \emph{predict} responses and 71\% to 87\% for \emph{explain} responses—with some exceptions in GPT-4-turbo. GPT-4-turbo demonstrated proficiency across all lessons with poorer performance relative to the other lessons for assessing \emph{predict} responses in \emph{Helping Students Manage Inequity} ($AUC$ = 0.17, $F_1$ = 0.45). The $AUC$ was low for scoring other lessons: \emph{predict} response in \emph{Exploring Implicit Bias} ($AUC$ = 0.43); \emph{explain} responses in \emph{Building Cultural Competence} ($AUC$ = 0.45), suggesting that the model performs poorly in distinguishing between correct and incorrect responses.

\begin{table*}[ht]
\centering
\caption{Comparison of absolute performance for \texttt{GPT-4o} across lessons for \textit{predict} and \textit{explain} open responses.}
\resizebox{\textwidth}{!}{%
\begin{tabular}{lccccc|ccccc}
\hline
\multirow{3}{*}{Lesson (Evaluation by \texttt{GPT-4o})} & \multicolumn{5}{c}{\multirow{2}{*}{\textit{Predict responses}}} & \multicolumn{5}{c}{\multirow{2}{*}{\textit{Explain responses}}} \\
                       & \multicolumn{5}{c}{}                      & \multicolumn{5}{c}{}                      \\ \cline{2-11} 
                       & Accuracy (\%)  & Precision  & Recall  & $AUC$  & $F_1$ score  & Accuracy (\%)  & Precision  & Recall  & $AUC$  & $F_1$ score  \\ \hline
\textit{Addressing Microaggressions}     & 79\%       & 0.80    & 0.71   & 0.78  & 0.75    & 76\%       & 0.92    & 0.59   & 0.77  & 0.72    \\
\textit{Avoiding Unconscious Assumptions}   & 75\%       & 0.81    & 0.75   & 0.75  & 0.78    & 76\%       & 0.65    & 0.74   & 0.44  & 0.69    \\
\textit{Building Cultural Competence}     & 91\%       & 0.86    & 0.89   & 0.45  & 0.88    & 71\%       & 0.83    & 0.74   & 0.72  & 0.78    \\
\textit{Exploring Implicit Bias}       & 71\%       & 0.98    & 0.69   & 0.79  & 0.81    & 74\%       & 0.85    & 0.66   & 0.76  & 0.75    \\
\textit{Helping Students Manage Inequity}   & 81\%       & 0.85    & 0.84   & 0.46  & 0.84    & 87\%       & 0.84    & 0.94   & 0.86  & 0.89    \\
\textit{Narrowing Opportunity Gaps}      & 82\%       & 0.93    & 0.78   & 0.84  & 0.85    & 85\%       & 0.54    & 0.78   & 0.42  & 0.64   \\\hline
\end{tabular}
}
\end{table*}
\begin{table*}[ht]
\caption{Comparison of absolute performance for \texttt{GPT-4-turbo} across lessons for \textit{predict} and \textit{explain} open responses.}
\resizebox{\textwidth}{!}{%
\centering
\begin{tabular}{lccccc|ccccc}
\hline
\multirow{3}{*}{Lesson (Evaluation by \texttt{GPT-4-turbo})} & \multicolumn{5}{c}{\multirow{2}{*}{\textit{Predict responses}}} & \multicolumn{5}{c}{\multirow{2}{*}{\textit{Explain responses}}} \\
                          & \multicolumn{5}{c}{}                      & \multicolumn{5}{c}{}                      \\ \cline{2-11} 
                          & Accuracy (\%)  & Precision  & Recall  & $AUC$  & $F_1$ score  & Accuracy (\%)  & Precision  & Recall  & $AUC$  & $F_1$ score  \\ \hline
\textit{Addressing Microaggressions}        & 77\%       & 0.85    & 0.60   & 0.76  & 0.70    & 75\%       & 0.92    & 0.56   & 0.75  & 0.70    \\
\textit{Avoiding Unconscious Assumptions}      & 70\%       & 0.57    & 0.67   & 0.47  & 0.61    & 78\%       & 0.98    & 0.72   & 0.84  & 0.83    \\
\textit{Building Cultural Competence}        & 79\%       & 0.86    & 0.78   & 0.43  & 0.82    & 59\%       & 0.64    & 0.56   & 0.45  & 0.60    \\
\textit{Exploring Implicit Bias}          & 85\%       & 0.87    & 0.83   & 0.43  & 0.85    & 79\%       & 0.85    & 0.76   & 0.79  & 0.80    \\
\textit{Helping Students Manage Inequity}      & 57\%       & 0.68    & 0.34   & 0.17  & 0.45    & 88\%       & 0.59    & 0.93   & 0.56  & 0.72    \\
\textit{Narrowing Opportunity Gaps}         & 75\%       & 1.00    & 0.60   & 0.80  & 0.75    & 73\%       & 0.88    & 0.63   & 0.76  & 0.73  \\\hline  
\end{tabular}
}
\end{table*}

\section{Discussion}
This study investigated differences in tutor learning across conditions that align with varying learning-by-doing activities (i.e., MCQ only, open response only, or both) and assessed the scalability of using generative AI for evaluating tutor responses. Several important insights emerged, which offer a comprehensive understanding of this present work.
\subsection{\textbf{No overall condition differences in learning outcomes.}}In summary, there was no main effect of learning-by-doing condition on posttest scores, or learning outcomes. However, there was a significant interaction between the condition and the lesson indicating some potential heterogeneity, meaning that the learning outcomes differed significantly by condition depending on the lesson the tutors completed. Two of the three significant pairwise comparisons (comparing lesson and condition) were quite close to the $p$-value threshold of 0.05 and with 18 such comparisons (3 comparisons for each of 6 lessons). So, perhaps these are chance occurrences. Nevertheless, 3 out of 18 significant differences are substantially more than the expected false positive rate of 1 in 20 comparisons and therefore worth further exploration. 
The pairwise comparison indicating that \textit{Open-response Only} is significantly better than \textit{Both} for \textit{Exploring Implicit Bias} is hard to interpret. It suggests that adding MCQ questions may harm learning in this lesson, and thus the \textit{MCQ Only} condition should be worse. However, the difference in learning outcomes between \textit{MCQ Only} and \textit{Open-response Only} was not significant ($p = .719$). Inspection of the \textit{Exploring Implicit Bias} did not reveal any substantial differences from other lessons in the nature of the MCQ or open-response questions. Overall, we have not found a consistent and generalizable explanation for the observed condition by lesson interaction. Perhaps the best explanation for this is random variability.
As a content-general conclusion, our evidence suggests that replacing open-response learning tasks with MCQ learning tasks produces generally equivalent learning outcomes. Certainly, we found no substantial evidence against MCQs as learning tasks. Indeed, we found no overall statistically reliable evidence for a decrease in learning outcomes due to the use of MCQs as learning tasks. In addition, there were no consistent trends across the six lessons in this direction. Indeed, the \textit{MCQ only} condition had the highest average posttest in two of the six lessons and the lowest only once. In contrast, the \textit{Open-response Only} condition had the highest posttest for two lessons, but the lowest for three.

\subsection{\textbf{The \textit{MCQ Only} condition requires less time.}}
There was no significant interaction between \textit{instruction time spent prior to posttest}, or time spent, and condition on learning outcomes. However, there was a significant interaction between the condition and the lesson, indicating that some lessons took longer to complete than others. In general, and not surprisingly, the \emph{MCQ Only} condition took the tutors the shortest time. However, the \emph{Both} condition took the longest time of the tutors, though not substantially longer than \emph{Open-response Only}, with the \emph{Both} condition taking the students less time than the sum of the open and MCQ conditions. Why did it not take learners longer to complete both forms of practice (\emph{Open-response} and \emph{MCQ}) compared to \emph{Open-response Only}? One possible explanation can arise from differences in the speed with which students in each condition complete the Follow-Up Instruction (Fig. \ref{fig:1_unnamed-6}). Conducting a \textit{ post hoc} ANOVA on follow-up instruction time revealed significant differences in follow-up instruction by condition, although that instruction included the same material in each condition (\emph{F}(2, 576) = 8.52, \emph{p} < .001). The mean comparisons indicated that learners in the \emph{Both} condition were faster to complete the follow-up instruction (\emph{M} = 0.155 minutes), followed by the \emph{MCQ Only} condition (\emph{M} = 0.239 minutes) and the \emph{Open-response Only} condition (\emph{M} = 0.347 minutes). These differences were statistically significant: the \emph{Both} condition was significantly faster in processing the follow-up instruction than the \emph{MCQ Only} condition, estimate = -0.47, \emph{SE} = 0.18, \emph{p} = .020, as well as the \emph{Open-response Only} condition, estimate = -0.78, \emph{SE} = 0.19, \emph{p} < .001.

\subsection{\textbf{LLMs demonstrate proficiency, but more research is needed for wide-scale assessment.}} Similar to \cite{lee2024applying}, we found the GPT models to be comparable to each other and overall demonstrated proficiency. However, GPT-4-turbo exhibited variability across lessons, with poorer performance in the \emph{Helping Students Manage Inequity} ($AUC$ = 0.17) and \emph{Exploring Implicit Bias} ($AUC$ = 0.43) lessons. The low $AUC$ scores in these cases suggest that the GPT-4-turbo struggled to classify responses in more complex or nuanced topics. This highlights the need for further refinement of LLMs to enhance their ability to assess open responses in content areas that require situational reasoning. In addition, more research is needed on the nuances within each lesson. The interrater reliability between the human graders was high for some lessons ($\kappa$ = 0.93), so it is necessary to revisit the human grade and annotation rubrics for the lessons and determine the sources of disagreement between the GPT models and humans (and between the raters) to better understand the results. This work adds to the recent literature on the potential use of LLMs for low stakes assessment tasks in different domain areas (Henkel et al. \cite{henkel2024can}), adding to the area of tutor training in advocacy skills.

\subsection{Limitations}
While this study used a posttest-only randomized experimental design, which offers several advantages, there are inherent limitations to this approach. One limitation is the lack of baseline data due to the absence of an analogous pretest scenario, which prevents measuring individual learning gains from pretest to posttest. This limitation makes it challenging to quantify the exact effect size of the lessons and to track individual progress. However, the primary purpose of this study was not to determine individual learning gain but to identify whether any of the learning-by-doing conditions--multiple choice questions, open-response questions, or both--led to better learning outcomes. By focusing on posttest results and employing random assignment to the condition, we were able to directly compare the relative learning effectiveness of each condition. In addition, the study aimed to find the most efficient lesson design by analyzing the time it took learners to complete each condition. Although the absence of pretest data can sometimes reduce statistical power, our sample size was sufficient to detect meaningful differences across conditions in time taken. Furthermore, our previous study \cite{thomas2023tutor}, which implemented the \textit{Both} condition with a counter-balanced pre-post design, demonstrated that our assessments are sufficient to detect significant pre-post learning gains. Finally, while the lack of baseline data can complicate the replication of the study in different contexts, the posttest only design was specifically chosen to minimize the risk of pretest sensitization, which can enhance the validity of the experimental results \cite{cook2007experimental, trochim2001research}. 

\section{Future Work and Conclusion}
We found evidence for learning efficiency benefits of multiple choice questions (MCQs) with feedback relative to the use of open-response questions with feedback or the combination of both. Practically, this result deserves particular attention given the widespread use of open-response questions in homework assignments as a practice task across content areas in college and K-12 education. This benefit of MCQs as learning tasks was demonstrated in a large sample of learners (n=235) and a variety of instructional lessons (6) and was revealed in generally equivalent learning outcomes achieved in significantly less time, corresponding to a 29\% practice time reduction compared to the open-response only and 35\% reduction compared a condition with both forms of instruction. Although these results favor the use of MCQs for instruction, they do not have any direct bearing on whether MCQs or open-response questions are better for assessing student learning. We used both in our posttest, and we intend to keep open-response questions to continue to do so to test that instructional practice with MCQ transfers to performance on open-ended questions. 

Our finding that MCQ practice transfers to open-ended performance is an important general result. The use of MCQs is often the target of criticism in assessment and instructional design. This criticism of MCQs as a shallow form of practice is perhaps more common when used for content that is less well defined, such as is the case for the learning goals of the tutoring lessons used in these studies (i.e., advocacy). Our evidence does not support such criticism and, further, provides evidence for the instructional benefits of MCQs relative to open-response questions in requiring less time for learners to reach the same outcome. Practically, this finding provides support for greater use, or at least greater exploration, of MCQs as practice tasks during instruction. Theoretically, this result has implications for the refinement of general frameworks for instructional design. In particular, the ICAP framework \cite{chi2014icap} makes a general prediction in the subject matter domain that constructive learning tasks, such as open-response questions, should produce better learning outcomes than active (but not constructive) learning tasks, such as feedback-based MCQs. Our learning outcome evidence is inconsistent with this prediction, especially as a generalization across learning content. 

We observed evidence of effect heterogeneity, with some lessons that included MCQ learning tasks yielding better learning outcomes, while others showed improved outcomes when MCQ tasks were excluded. This content-treatment interaction has been found and well explained in previous research (e.g., \cite{rachatasumrit2023content}); however, in this case, the explanation is not clear, and we suggest future work to probe whether there is a replicable finding here and, if so, what theory might explain it. Finally, given how prevalent open-response questions are as learning tasks in homework assignments in school and college education, we recommend further investigation of the potential for more efficient and equally effective learning from the use of multiple-choice questions as learning tasks. 

\begin{acks}
 This work was made possible with the support of the Learning Engineering Virtual Institute. The opinions, findings, and conclusions expressed in this material are those of the authors.
\end{acks}

\bibliographystyle{ACM-Reference-Format}
\bibliography{main}

\appendix

\section{Digital Appendix}
All analysis code, study materials, and log data references can be found in the study's supplementary GitHub repository:\\
\url{https://github.com/CMU-PLUS/LAK2025-Advocacy}

\end{document}